\documentclass[rnaas]{aastex63}

\newcommand{\swift}{\textit{Swift}}

\submitjournal{RNAAS}

\shorttitle{ROFL}
\shortauthors{Tohuvavohu}
\graphicspath{{./}{figures/}}
\begin{document}

\title{Rotation Optimized Filter for Longevity (ROFL): Increasing the lifetime of Swift/UVOT simply}

\correspondingauthor{Aaron Tohuvavohu}
\email{tohuvavohu@astro.utoronto.ca}

\author[0000-0002-2810-8764]{Aaron Tohuvavohu}
\affiliation{Department of Astronomy \& Astrophysics \\
University of Toronto \\
Toronto, ON, Canada}

%% Note that the \and command from previous versions of AASTeX is now
%% depreciated in this version as it is no longer necessary. AASTeX 
%% automatically takes care of all commas and "and"s betIen authors names.

%% AASTeX 6.3 has the new \collaboration and \nocollaboration commands to
%% provide the collaboration status of a group of authors. These commands 
%% can be used either before or after the list of corresponding authors. The
%% argument for \collaboration is the collaboration identifier. Authors are
%% encouraged to surround collaboration identifiers with ()s. The 
%% \nocollaboration command takes no argument and exists to indicate that
%% the nearby authors are not part of surrounding collaborations.

%% Mark off the abstract in the ``abstract'' environment. 
\begin{abstract}

I demonstrate that a very simple and safe change to the planning software filter assignment algorithm for the Ultraviolet/Optical Telescope (UVOT) onboard the Neil Gehrels Swift Observatory can reduce the number of filter wheel rotations by $>10\%$, and its adoption is thus likely to significantly extend the usable lifetime of the UVOT instrument. 

\end{abstract}

%% Keywords should appear after the \end{abstract} command. 
%% See the online documentation for the full list of available subject
%% keywords and the rules for their use.
%\keywords{space telescopes --- ultraviolet astono
%miscellaneous --- catalogs --- surveys}

%% From the front matter, I move on to the body of the paper.
%% Sections are demarcated by \section and \subsection, respectively.
%% Observe the use of the LaTeX \label
%% command after the \subsection to give a symbolic KEY to the
%% subsection for cross-referencing in a \ref command.
%% You can use LaTeX's \ref and \label commands to keep track of
%% cross-references to sections, equations, tables, and figures.
%% That way, if you change the order of any elements, LaTeX will
%% automatically renumber them.
%%
%% I recommend that authors also use the natbib \citep
%% and \citet commands to identify citations.  The citations are
%% tied to the reference list via symbolic KEYs. The KEY corresponds
%% to the KEY in the \bibitem in the reference list below. 

\section{Introduction} \label{sec:intro}
The Ultra-Violet Optical Telescope (UVOT; \citet{UVOT}) onboard the Neil Gehrels \swift\ Obsevatory \citep{swift2004} is a workhorse instrument of the time-domain community, and the only instrument regularly performing transient science and fast follow-up in the UV. The UVOT has a photon counting detector with a bandpass from 170-700 nm, covering a 17'x17' field-of-view. In the optical path is an 11-position filter wheel, carrying 7 filters, 2 grisms, a focal expander (magnifier), and a blocking plate. The filter wheel rotates on a stub axle, and is driven by a uni-directional stepper motor. The filter wheel was rated for 50,000 revolutions in its design, and has substantially surpassed this number to date. The UVOT filter wheel is the only regularly moving mechanism in any of the \swift\ instruments, and is likely to be the main lifetime limiting component for UVOT. 

\section{Filter-of-the-Day}
For this reason, limiting and optimizing the number of rotations made by the filter wheel is paramount. The method adopted at the beginning of the \swift\ mission, and used still today, is the `Filter of the Day' (FOTD)\footnote{\href{https://www.swift.psu.edu/secure/toop/uvot_filter_rotation_info.htm}{https://www.swift.psu.edu/secure/toop/uvot\_filter\_rotation\_info.htm}} scheme: whereby each day is assigned one of the four near-UV filters, in a repeating cycle, and all observations not requiring a particular mode or filter set from UVOT (ie observations for which the main science is being performed with the XRT) are set in this filter. The near-UV (and not optical) filters are chosen so as to ensure there is still a unique UV contribution to the observation. Such an arrangement is designed to have as many consecutive observations be scheduled in the same filter as possible to eliminate the need for movement of the filter wheel. 

At the beginning of the \swift\ mission the vast majority of observations were XRT driven, and the science cases required no specific UVOT filters or modes, thus allowing for a substantial number of consecutive observations with UVOT set in the FOTD, and proving the efficacy of this approach to reducing the filter wheel rotations. However, over the course of the \swift\ mission the UVOT has become a high-demand instrument for the community, for science ranging from comet observations in the inner solar system \citep{comets}, near-UV exoplanet transits \citep{exoplanet}, the mapping of star formation histories of nearby galaxies \citep{hagen2017}, and most predominantly, providing early-time UV light curves for $\sim1000$ supernovae to date \citep{sousa}, along with myriad other transients (eg the first UV detection of a kilonova \cite{evans2017}). The success of the community in fully utilizing UVOT, as well as increased demand on \swift\ overall, has changed the typical \swift\ schedule. While the FOTD technique is still used when appropriate, the majority of science programs require specific UVOT filters and modes, and thus the number of consecutive observations in FOTD are significantly reduced. In addition the \swift\ schedule has grown denser, with more individual observations performed on average per day each year ($>100$ per day in 2019).

\section{An Alternative}
Under these substantially different scheduling conditions for UVOT, and in the context of larger advancements in \swift\ scheduling \citep{acat,deich}, I examined the efficacy of the FOTD scheme for reducing filter wheel rotations and found that it substantially under-performed with respect to a simple alternative:
\begin{verbatim}
    
For any observation without specific filters required for the science 
(ie for which the FOTD would normally be assigned):
If the last position of the filter wheel was on a UV filter:
    Leave the filter wheel in this position
Else:
    Rotate to the nearest (directional) UV filter.
\end{verbatim}

I rebuilt the \swift\ observing schedules for all of 2018 and 2019 using this schema, and find a reduction of $>18,000$ filter wheel movements per year as compared to the FOTD scheme. This corresponds to a $>10\%$ reduction in the number of rotations performed, and thus likely a similar extension to the lifetime of the filter wheel. This new scheme, which I call the Rotation Optimized Filter for Longevity (ROFL), naturally entails a different distribution in the use of the UV filters, which I show in Figure 1:

\begin{figure}[h!]
    \centering
    \includegraphics[width=0.7\textwidth]{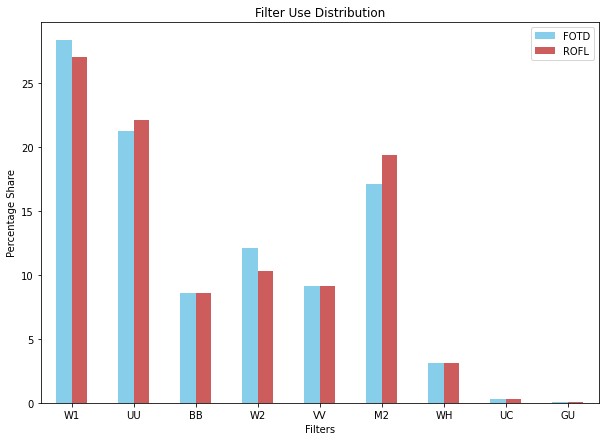}
    \caption{The filter use distribution of all \swift/UVOT observations scheduled in 2019. Shown for both the FOTD scheme, and the proposed ROFL scheme. ROFL results in a slight increase in the number of observations taken with the $u$ and $uvm2$ filters, and a corresponding decrease in the use of the $uvw1$ and $uvw2$ filters.}
    \label{fig:my_label}
\end{figure}

Given the simplicity of this scheme and the large impact it can have on reducing the wear on the UVOT filter wheel, and therefore likely increasing the functional lifetime of the instrument, I recommend that it be implemented within the \swift\ planning software.

%% For this sample I use BibTeX plus aasjournals.bst to generate the
%% the bibliography. The sample63.bib file was populated from ADS. To
%% get the citations to show in the compiled file do the following:
%%
%% pdflatex sample63.tex
%% bibtext sample63
%% pdflatex sample63.tex
%% pdflatex sample63.tex

\bibliography{main}{}
\bibliographystyle{aasjournal}

%% This command is needed to show the entire author+affiliation list when
%% the collaboration and author truncation commands are used.  It has to
%% go at the end of the manuscript.
%\allauthors

%% Include this line if you are using the \added, \replaced, \deleted
%% commands to see a summary list of all changes at the end of the article.
%\listofchanges

\end{document}